\documentclass[fleqn,10pt,twocolumn]{AROB-ISBC-SWARM23}

\usepackage{here}
\newcommand{\figurename}{Fig.}

\def\fgref#1{\figurename\ref{#1}}

\usepackage{tabularx}
\usepackage{subfigure}
\usepackage{comment}

\title{Deep-learning models in medical image analysis:\\
       Detection of esophagitis from the Kvasir Dataset}

\author{Kyoka Yoshioka${}^{1\dagger}$, Kensuke Tanioka${}^{2}$, Satoru Hiwa${}^{2}$ and Tomoyuki Hiroyasu${}^{2}$}
\speaker{Kyoka Yoshioka}

\affils{${}^{1}$Graduate School of Life and Medical Sciences, Doshisha University, Kyoto, Japan\\
${}^{2}$Department of Biomedical Sciences and Informatics, Doshisha University, Kyoto, Japan\\
(Tel: +81-774-65-6020; E-mail: tomo@is.doshisha.ac.jp)\\
}
\abstract{%
Early detection of esophagitis is important because this condition can progress to cancer if left untreated. However, the accuracies of different deep learning models in detecting esophagitis have yet to be compared. Thus, this study aimed to compare the accuracies of convolutional neural network models (GoogLeNet, ResNet-50, MobileNet V2, and MobileNet V3) in detecting esophagitis from the open Kvasir dataset of endoscopic images. Results showed that among the models, GoogLeNet achieved the highest F1-scores. Based on the average of true positive rate, MobileNet V3 predicted esophagitis more confidently than the other models. The results obtained using the models were also compared with those obtained using SHapley Additive exPlanations and Gradient-weighted Class Activation Mapping.
}


\begin{document}

\maketitle


\section{Introduction}
With the development of artificial intelligence (AI), several studies have focused on the application of this technology in the medical field. In gastroenterology, AI is used to detect inflammation, polyps, and stomach cancer and develop systems that can automatically determine the severity of symptoms \cite{chen2018accurate} \cite{hirasawa2018application} \cite{guimaraes2020deep} \cite{zhang2020diagnosing}.
AI models are expected to improve diagnostic accuracy and reduce medical costs by preventing misdiagnosis by humans.

Various deep learning and AI models, including deep learning convolutional neural network (CNN) models, have been proposed and used for medical image recognition and analysis.
However, these models differ in accuracy, and comparing this aspect is important to identify which model is suitable for a specific application in endoscopic imaging.

The z-line is an anatomic landmark located posterior to the stomach and esophagus. Esophagitis is an inflammation of the esophagus that appears as a break in the esophageal mucosa relative to the z-line \cite{pogorelov2017kvasir}. The z-line and esophagitis can be described as normal and diseased conditions, respectively. Early detection of esophagitis is necessary because this condition can cause complications (e.g., esophageal ulcer, bleeding, and stricture) and progress to cancer if left untreated. Therefore, distinguishing between the z-line and esophagitis is necessary. However, this procedure is difficult \cite{cogan2019mapgi}. In addition, the accuracies of various models in detecting esophagitis have yet to be compared.

Thus, this study aimed to compare the accuracies of several CNN models, including GoogLeNet \cite{szegedy2015going}, ResNet-50 \cite{he2016deep}, MobileNet V2 \cite{sandler2018mobilenetv2}, and MobileNet V3 \cite{howard2019searching}, in identifying z-lines and esophagitis in endoscopic images from the open Kvasir dataset. These models have received considerable attention in recent years after winning in the ImageNet Large Scale Visual Recognition Challenge (ILSVRC), a competition using a large image recognition dataset. The results obtained by the four CNN models were compared. The training models were also compared with the explainable artificial intelligence (XAI) methods  Gradient-weighted Class Activation Mapping (Grad-CAM) \cite{selvaraju2017grad} and SHapley Additive exPlanations (SHAP) \cite{lundberg2017unified}.

\section{Deep learning in medical image analysis}
\subsection{Typical architecture for image classification}
CNN is a deep learning method specialized for image recognition. It is widely used for identifying lesion sites in medical images. It combines a convolutional layer with a pooling layer and finally iterates through all the combined layers to generate the results. In this study, we compared the results of different CNN models used for site identification in medical images. The CNN models used included GoogleNet and ResNet, the successive winning models of ILSVRC, and MobileNet V2 and MobileNet V3, which have attracted considerable attention in recent years because of their small computational and memory.

\subsubsection{GoogLeNet}
GoogLeNet was the winning model at ILSVRC in 2014
The model consists of an Inception module, 1$\times$1 convolution, auxiliary loss, and global average pooling. GoogLeNet can be multi-layered using the Inception module, but 1$\times$1 convolution is performed before each convolution calculation to reduce dimensionality resulting from the large number of parameters. The Inception module helps process data using multiple filters in parallel. The fully connected layer is removed to increase the width and depth of the network, average pooling is used instead of the fully connected layer to avoid gradient loss, and class classification is performed on sub-networks branched from the middle of the network by auxiliary loss \cite{szegedy2015going}.

\subsubsection{ResNet}
ResNet was the winning model at the ILSVRC in 2015. The problem of learning not progressing due to gradient loss and degradation problems was solved using a method called Residual Block, which uses 152 very deep layers to solve the problem. The key features of this model are residual block and batch normalization using shortcut connection. ResNet has several models with different layer depths. ResNet-50 shows higher accuracy than GoogLeNet in ImageNet classification \cite{he2016deep}. However, ResNet-50 requires about twice as many parameters as GoogLeNet.

\subsubsection{MobileNet V2}
MobileNet is a small computationally and memory model that can adjust the trade-off between accuracy and computational load. Depthwise separable convolution decomposes the convolution layer into depthwise and pointwise convolution for computation. This mechanism reduces the computation cost.
Furthermore, V2 introduces expand/projection layers and inverted residual blocks. Expand/projection layers rapidly increase or decrease the number of channels. MobileNet V2 achieves comparable accuracy to GoogLeNet and ResNet-50 in ImageNet classification while significantly reducing the number of parameters \cite{sandler2018mobilenetv2}.

\subsubsection{MobileNet V3}
MobileNet V3 is an improved version of MobileNet V2, introducing a squeeze-and-excite structure (SE-block) in the inverted residual block, one of the features of MobileNet V2. SE-block improves the expressiveness of the model by weighting information in the channel direction \cite{hu2018squeeze}.
Compared with V2, MobileNet V3 shows more accurate ImageNet classification while shortening total inference time \cite{howard2019searching}.

\subsection{Explainable AI (XAI)}
The CNN models were compared with XAI methods Grad-CAM and SHAP.
The Discussion section explains the results obtained using these techniques.

\subsubsection{Grad CAM}
Grad-CAM displays a color map of the area the CNN is gazing at for classification \cite{selvaraju2017grad}.
It is based on the fact that variables with large gradients in the output values of the predicted class are essential for classification prediction. The gradient of each input image pixel with respect to the output value of the prediction class in the last convolution layer is used.

\subsubsection{SHAP}
SHAP calculates, for each predicted value, how each characteristic variable affects that prediction \cite{lundberg2017unified}. This analysis allows us to visualize the impact of an increase or decrease in the value of a given characteristic variable.

\section{Materials and Methods}
CNN models GoogLeNet, ResNet-50, MobileNet V2, and MobileNet V3 were employed to detect esophagitis from the open Kvasir dataset of endoscopic images, and their results were compared.

\subsection{Kvasir dataset}
The Kvasir dataset is a collection of endoscopic images of the gastrointestinal tract. It was annotated and validated by certified endoscopists. The dataset was made available in the fall of 2017 through the Medical Multimedia Challenge provided by MediaEval. It includes anatomical landmarks (pylorus, z-line, and cecum), disease states (esophagitis, ulcerative colitis, and polyps), and medical procedures (dyed lifted polyps and dyed resection margins). The resolution of the images from the Kvasir dataset with these eight classes varies from 720$\times$576 pixels to 1920$\times$1072 pixels. Each image has a different shooting angle, resolution, brightness, magnification, and center point.

\subsection{Prepossessing}
Image prepossessing was performed before training the models.
Edge artifacts and annotations that interfere with learning during the analysis of medical images were removed. A mask image was created, where pixels with luminance values below a certain threshold were set to 0. The opening process was applied to the mask image to remove the annotations. The image was cropped using this final mask image to obtain the target area. This process was performed on all data.

Each image in the dataset has a different resolution. All images were resized to 224$\times$224 pixels by bilinear completion and optimized for deep learning input. In addition to these processes, data augmentation was performed on the data used for learning. We applied two types of data augmentation: horizontal and vertical flip.

\subsection{Cross Validation}
A total of 1000 image data sets containing z-lines and esophagitis were partitioned into test, training, and validation data. First, 25\% (n = 250) of the total data were randomly selected to generate test data. Of the remaining data (75\%, n = 750), 50\% (n = 500) was used for training and 25\% (n = 250) for validation.

The inner loop consisted of training and validation data. The model was trained using the training data, and parameters such as the optimal number of epochs were determined using the validation data. Thus, four training models were generated. The test data of each model were evaluated, and the average of discrimination accuracy of the four times was used as the evaluation value of the CNN model. The test, training, and validation data were each partitioned to maintain the class proportions.

\subsection{CNN models}
PyTorch was used for the implementation of GoogLeNet, ResNet-50, MobileNet V2, and MobileNet V3. The initial values of all model parameters were pre-trained by ImageNet, and the models were trained by fine tuning.

For all models, the Adam optimizer was used for training. The batch size was five, and the maximum number of epochs was 100.
The cross-entropy error shown in equation (\ref{eq:cross loss}) was used as the loss function.

\begin{eqnarray}
E(\rm{x})
&=& -\sum_{n=1}^{N}\sum_{k=1}^{K}{d}_{nk}\log{y}_{k}(\rm{x}_{n}; \rm{w})
\label{eq:cross loss}
\end{eqnarray}

\subsection{Evaluation Function}
Five evaluation indices were used in this experiment: accuracy, precision, recall, specificity, and F1-score. These metrics were calculated using the confusion matrix shown in Table \ref{tb:confusion_matrix}.

\begin{table}[H]
\label{confusion_matrix}
\centering
\caption{\label{tb:confusion_matrix} Confusion matrix for a two-class problem}
\begin{tabular}{l|l|l}
\hline
& \begin{tabular}[c]{@{}l@{}}Predicted Class\\ (Positive Class)\end{tabular} & \begin{tabular}[c]{@{}l@{}}Predicted Class\\ (Negative Class)\end{tabular} \\ \hline\hline
\begin{tabular}[c]{@{}l@{}}Actual Class\\ (Positive Class)\end{tabular}  & True Positive                                                          & False Negative                                                         \\
\begin{tabular}[c]{@{}l@{}}Actual Class \\ (Negative Class)\end{tabular} & False Positive                                                         & True Negative                                                          \\ \hline\hline
\end{tabular}
\end{table}

In this experiment, the z-line and esophagitis were judged as the negative and positive classes, respectively. In other words, data judged to be esophagitis and z-line by the learning model were designated true positive (TP) and false negative (FN), respectively. Meanwhile, data determined to be esophagitis and z-line by the training model were designated false positive (FP) and true negative (TN), respectively. Based on the values of TP, FP, TN, and FN obtained from the confusion matrix, the accuracy, precision, recall, specificity, and F1-score of the models were calculated using Equations(\ref{eq:Accuracy}) to (\ref{eq:F1score}).

\begin{equation}
Accuracy=\frac{TP+TN}{TP+FP+FN+TN}
\label{eq:Accuracy}
\end{equation}

\begin{equation}
Precision=\frac{TP}{TP+FP}
\label{eq:Precision}
\end{equation}

\begin{equation}
Recall=\frac{TP}{TP+FN}
\label{eq:Recall}
\end{equation}

\begin{equation}
Specificity=\frac{TN}{TN+FP}
\label{eq:Specificity}
\end{equation}

\begin{equation}
F1 \: score=\frac{2TN}{2TP+FP+FN}
\label{eq:F1score}
\end{equation}

\section{Results and Discussions}
\subsection{Performance comparison between different architecture}
The evaluation indices obtained from the experiments are shown in Table \ref{tb:performance comparison}.

\begin{table}[h]
\caption{\label{tb:performance comparison}Performance comparison between different architecture}
\begin{center}
\begin{tabularx}{\linewidth}{lllllll}\hline
Model & \begin{tabular}[c]{@{}l@{}}ACC\end{tabular}& \begin{tabular}[c]{@{}l@{}}PREC\end{tabular}& \begin{tabular}[c]{@{}l@{}}REC\end{tabular} & \begin{tabular}[c]{@{}l@{}}SPEC\end{tabular}& \begin{tabular}[c]{@{}l@{}}F1\end{tabular}     \\ \hline\hline
GoogLeNet & \textbf{0.846} & 0.859 & \textbf{0.830} & 0.862 & \textbf{0.843}  \\
MobileNet V3 & 0.842 & \textbf{0.901} & 0.776 & \textbf{0.908} & 0.831 \\
ResNet-50 & 0.833 & 0.865 & 0.792 & 0.874 & 0.826 \\
MobileNet V2 & 0.830 & 0.852 & 0.800 & 0.860 & 0.825 \\
\hline
\end{tabularx}
\end{center}
\end{table}

The F1-score results in Table \ref{tb:performance comparison} show that GoogLeNet was the best among the four models.
In other words, GoogLeNet was more reliable in predicting esophagitis than the other models. Meanwhile, MobileNet V3 showed the highest precision and specificity. In other words, MobileNet V3 was the most accurate among the tested models for z-line prediction. From a medical point of view, an ideal model should be likely to distinguish esophagitis with severe symptoms from the z-line.

The average of TP rate were 0.950, 0.923, 0.892, and 0.841 for MobileNet V3, MobileNet V2, GoogLeNet, and ResNet-50, respectively. MobileNet V3 predicted esophagitis with more confidence than the other models.

\subsection{GoogLeNet analysis}
Grad-CAM and SHAP were applied to the learned model, and what kind of the model was created was discussed.

\begin{figure*}[t]
\begin{center}
\includegraphics[width=14.3cm]{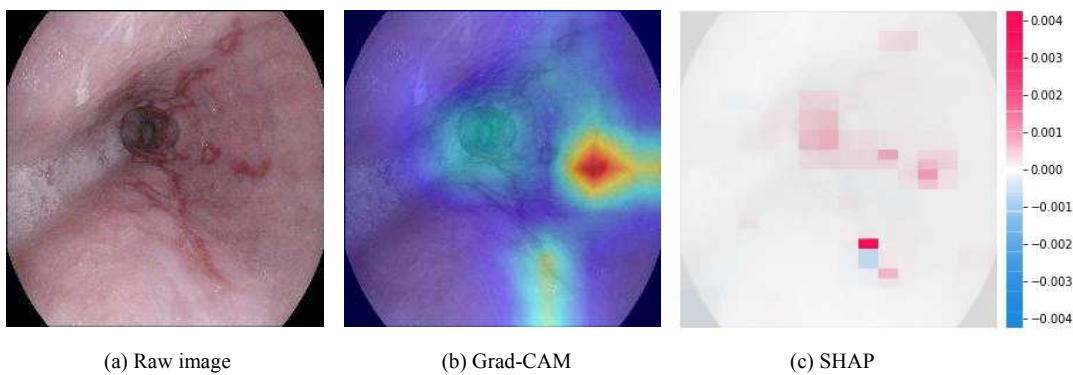}
\caption{\label{True Positive Pattern}True Positive Pattern}
\end{center}
\end{figure*}

\fgref{True Positive Pattern} shows an example of the image results in the case of TP predicted by GoogLeNet. In the Grad-CAM results, red indicates the most potent activation, and blue indicates the weakest activation. In the SHAP results, the SHAP values of the patches were computed and rendered in a color map: a positive SHHAP value (red) indicates that the class is supported. By contrast, a negative SHAP value (blue) indicates that the class is rejected.

Tearing the esophageal mucosa against the z-line is a feature of esophagitis. According to \fgref{True Positive Pattern}, the results of Grad-CAM and SHAP showed that the learned model of GoogLeNet can makes predictions focusing on the clinically significant aspects of esophagitis images. The GoogLeNet model learned the findings that are important for diagnosing esophagitis. Comparison results showed that SHAP captured the location of multiple mucosal tears in the image more accurately than Grad-CAM.

\begin{figure*}[t]
\begin{center}
\includegraphics[width=14.3cm]{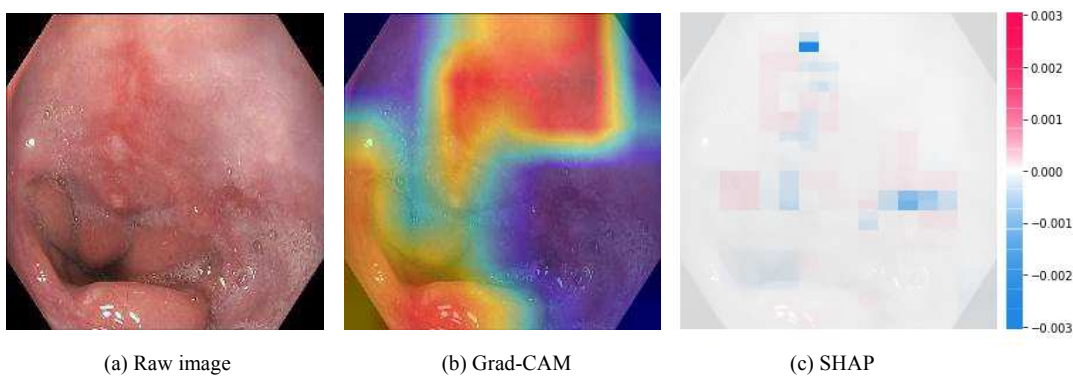}
\caption{\label{False Negative Pattern}False Negative Pattern}
\end{center}
\end{figure*}

\fgref{False Negative Pattern} shows the results of applying Grad-CAM and SHAP in the FN case. The following can be observed from the results of Grad-CAM and SHAP for \fgref{False Negative Pattern}, respectively. In the Grad-CAM results, most areas in the image are shown as activated regions.
Areas that provide the basis for the prediction are difficult to identify because of the gradient saturation in the Grad-CAM calculation. In the SHAP results, the inflammatory areas of the input image are indicated by blue pixels. Blue pixels indicate features that have a negative contribution to the prediction. In other words, although the model incorrectly identified esophagitis as a z-line, the model recognized that areas in the image negatively contributed to the z-line decision.

\subsection{MobileNet V3 analysis}
\begin{figure*}[tb]
 \begin{center}
  \subfigure{
   \includegraphics[width=.63\columnwidth]{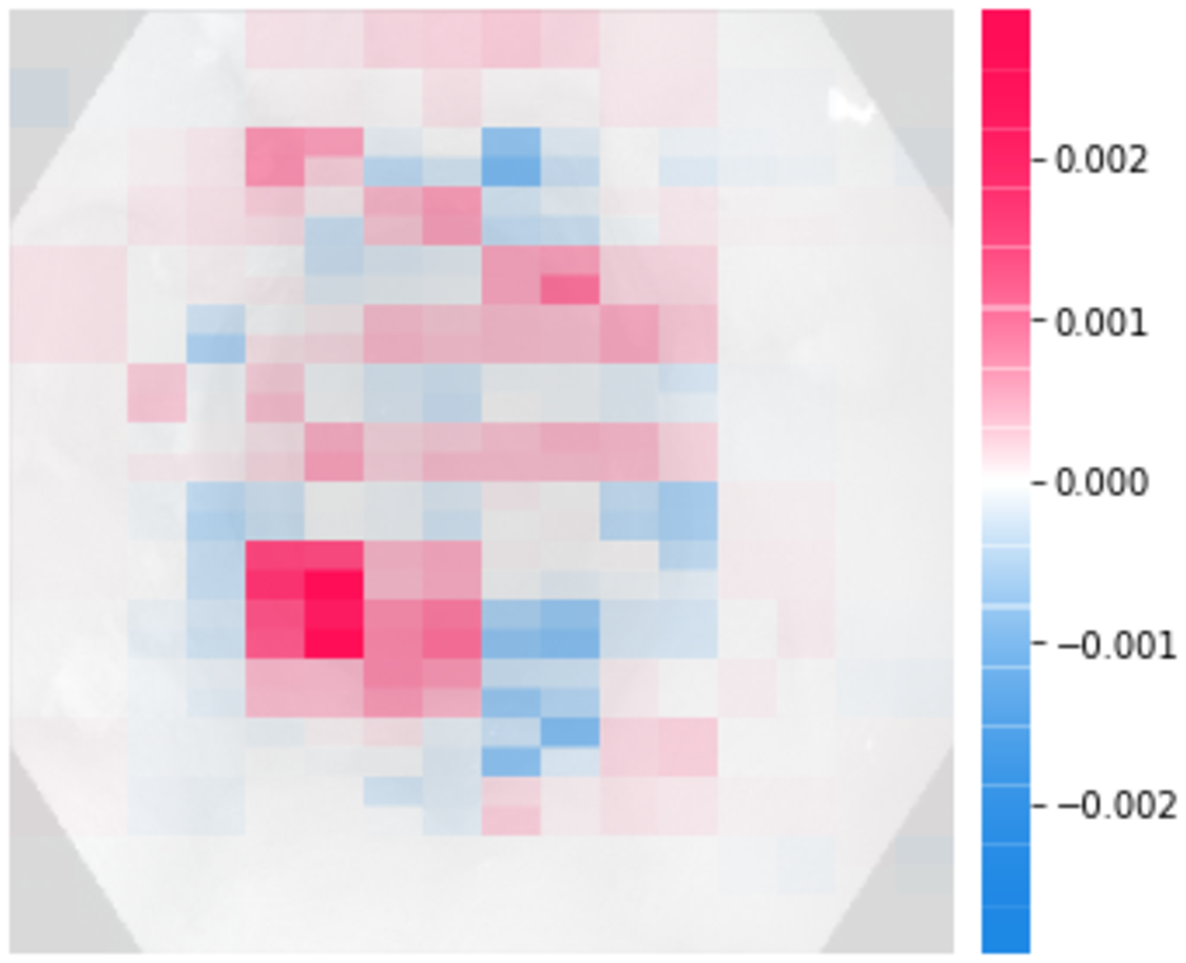}
  }~
  \subfigure{
   \includegraphics[width=.63\columnwidth]{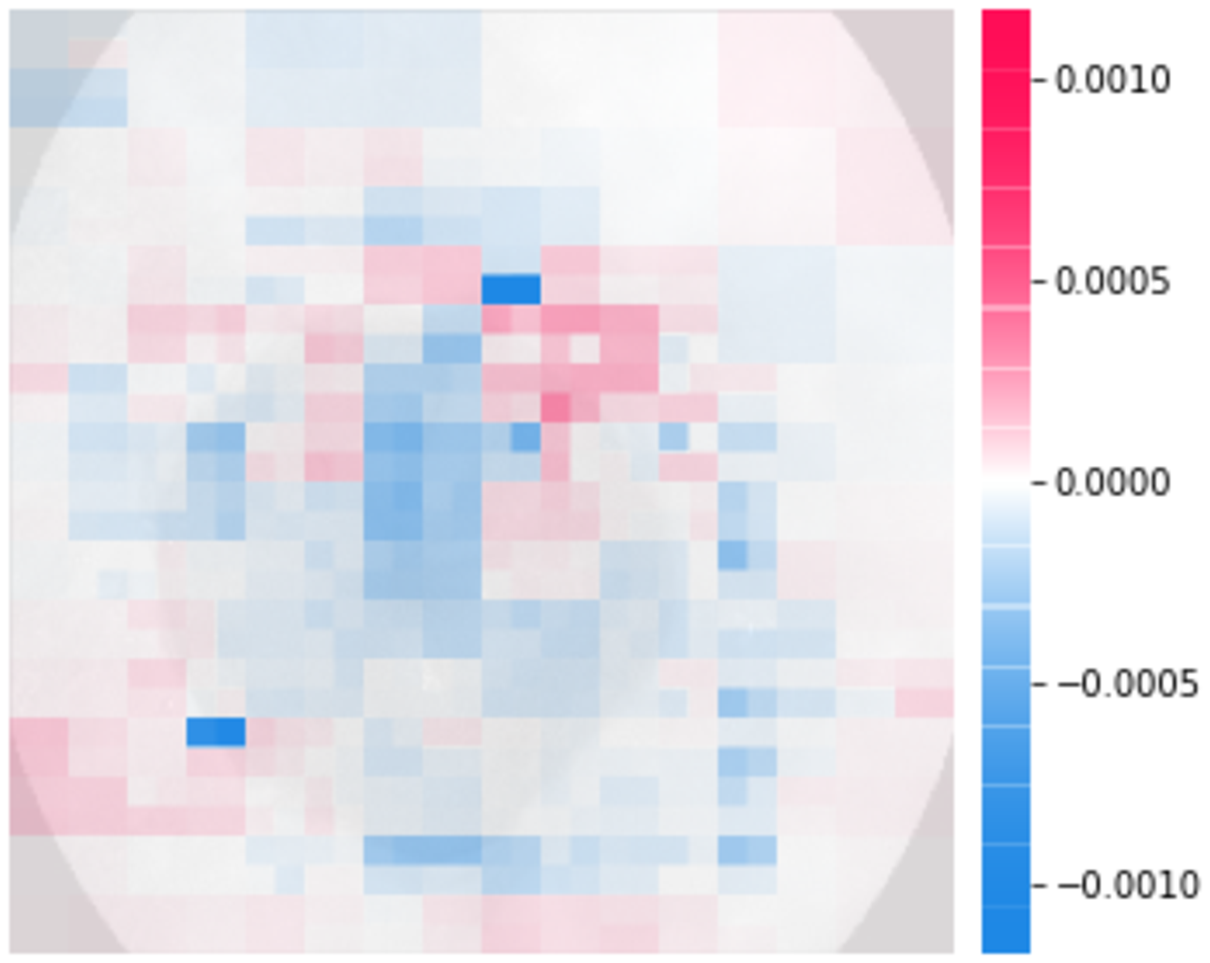}
  }
  \caption{\label{Prob images}First image predicted positive with 1.000 probability, and second image predicted positive with 0.524 probability.} 
 \end{center}
\end{figure*}

One hundred images were determined to be TP in the MobileNet V3 model. The SHAP results for the images judged to have the highest and lowest probabilities of being esophagitis are shown in \fgref{Prob images}.

As shown in \fgref{Prob images}, in cases with a high prediction probability, some features may have a negative contribution to the prediction. Many features showing negative contributions can be identified in the images with low prediction probability for \fgref{Prob images}. In this case, the prediction probability may be low.

\section{Conclusions}
We compared the accuracies of CNN models, including GoogLeNet, ResNet-50, MobileNet V2, and MobileNet V3, in identifying z-line and esophagitis in endoscopic images from the open Kvasir dataset. Among the four models, GoogLeNet had the highest F1-score, and MobileNet V3 had the highest average TP rate. These results suggest that GoogLeNet performs better than state-of-the-art CNN models in medical image recognition. In addition, MoblieNet V3 is a cost-effective model because of its low memory and short training time. Each model was analyzed and compared with Grad-CAM, and SHAP.
Other models, datasets, and model analyses are warranted for verification.


\bibliographystyle{unsrt}
\bibliography{AROB-ISBC-SWARM_kyyoshioka}

\end{document}